\documentclass{Interspeech}

\interspeechcameraready 

\newcommand\firstpara[1]{\noindent\textbf{#1}\,}
\newcommand\para[1]{\noindent\textbf{#1}\,}

\title{DualCodec: A Low-Frame-Rate, Semantically-Enhanced Neural Audio Codec for Speech Generation}

\author[affiliation={1}]{Jiaqi}{Li}
\author[affiliation={2}]{Xiaolong}{Lin}
\author[affiliation={1}]{Zhekai}{Li}
\author[affiliation={2}]{Shixi}{Huang}
\author[affiliation={1}]{Yuancheng}{Wang}
\author[affiliation={1}]{Chaoren}{Wang}
\author[affiliation={2}]{Zhenpeng}{Zhan}
\author[affiliation={1}]{Zhizheng}{Wu}

\affiliation{School of Data Science}{The Chinese University of Hong Kong, Shenzhen}{China}
\affiliation{}{Baidu Inc., Shenzhen}{China}
\email{jiaqili3@link.cuhk.edu.cn}

\usepackage{comment}
\usepackage{amssymb}
\usepackage{amsmath}
\usepackage{amsmath,graphicx}
\usepackage{multirow}
\usepackage{makecell}
\usepackage{booktabs}
\usepackage{subcaption}
\usepackage{pifont} 
\usepackage{tikz}

\usepackage{hyperref}

\begin{document}
\maketitle
\begin{abstract}

Neural audio codecs form the foundational building blocks for language model (LM)-based speech generation. Typically, there is a trade-off between frame rate and audio quality. This study introduces a low-frame-rate, semantically enhanced codec model. Existing approaches distill semantically rich self-supervised (SSL) representations into the first-layer codec tokens. This work proposes \textbf{\textit{DualCodec}}, a dual-stream encoding approach that integrates SSL and waveform representations within an end-to-end codec framework. In this setting, DualCodec enhances the semantic information in the first-layer codec and enables the codec system to maintain high audio quality while operating at a low frame rate. Note that a low-frame-rate codec improves the efficiency of speech generation. Experimental results on audio codec and speech generation tasks confirm the effectiveness of the proposed DualCodec compared to state-of-the-art codec systems, such as Mimi Codec, SpeechTokenizer, DAC, and Encodec.
Demos are available at:
\url{https://dualcodec.github.io}, code is available at: \url{https://github.com/jiaqili3/DualCodec}
\end{abstract}

\section{Introduction}
Neural audio codec is a technique to compress audio signals into a series of discrete codes for efficient data storage and transmission~\cite{ts3codec,encodec,soundstream}.
Recently, they are more frequently utilized as the tokenizers and de-tokenizers of speech language models (TTSs)~\cite{valle,audioLM2023,valle2,soundstorm2023,maskgct}. 
Inspired by the success of large language models, 
TTSs model discrete speech tokens in a language modeling framework,  and have shown impressive results in zero-shot text-to-speech (TTS)~\cite{valle}.
In a typical TTS framework like VALL-E~\cite{valle}, a neural audio codec such as Encodec~\cite{encodec}
encodes waveform signal into hierarchical discrete speech tokens using residual vector quantization (RVQ).
The first-layer codebook tokens (RVQ-1) are predicted by an autoregressive (AR) language model conditioned on text, and the remaining codebook layers (RVQ-rest) are predicted via a non-autoregressive (NAR) language model.
Then, the codec decoder converts the speech tokens into audio.

Although this SLM TTS framework has impressive zero-shot TTS capabilities, 
it still suffers from problems like inaccurate speech \textit{content}, limited speech generation \textit{quality}, and slow inference \textit{speed}~\cite{speechtokenizer,codecinvestigation,moshi}. 
These three problems are closely related to the speech tokens.
Motivated by recent works on improving each of these aspects~\cite{speechtokenizer,moshi,dac}, we summarize important design principles for a practical speech generation-oriented neural audio codec:

\begin{itemize}
    \item \textbf{Semantic enhancement}: Self-supervised (SSL) speech features 
     contain rich pronunciation and semantic information~\cite{bertphone}. Previous work SpeechTokenizer~\cite{speechtokenizer} distilled SSL feature into a neural audio codec and found this benefits TTS intelligibility and stability. 
    \item \textbf{Low frame rate}: A low frame rate codec decreases the token length, reducing the resources to train and inference TTS. 
    \item \textbf{Audio quality}: A high codec reconstruction quality is essential to TTS's audio quality~\cite{codecinvestigation}. 
    This becomes challenging for low-frame-rate audio codecs operating at a low bitrate.
\end{itemize}

\begin{figure*}
    \centering
    \vspace{-6mm}
    \includegraphics[width=0.9\linewidth]{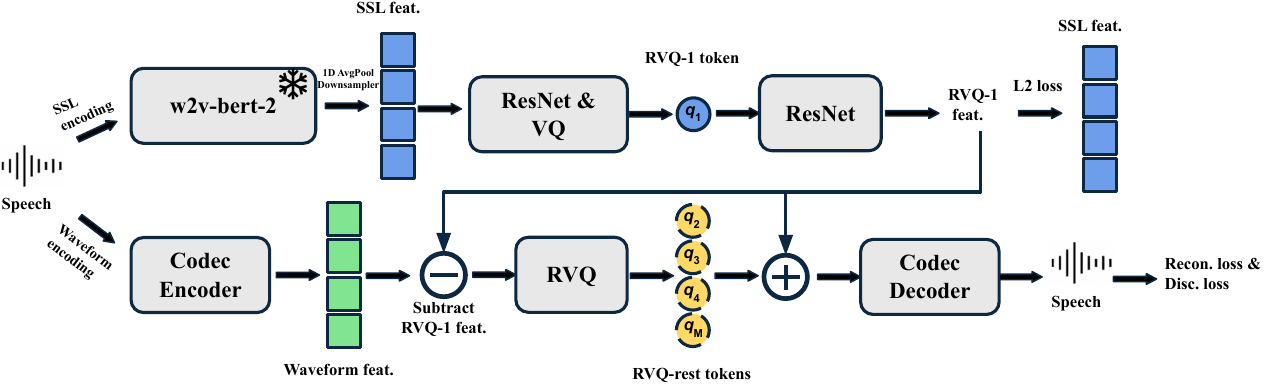}
    \vspace{-2mm}
    \caption{The dual encoding method for neural audio codecs. 
    The upper stream is SSL encoding, and the lower stream is waveform encoding.
    Given a speech input, the SSL feature is obtained from a pretrained w2v-bert-2~\cite{seamless} model and then encoded as the codec's first-layer token (RVQ-1).
    The remaining RVQ layers encodes the residual between the waveform feature and the RVQ-1 feature, and outputs audio. 
    The framework is trained end-to-end requiring an additional L2 SSL feature loss in addition to codec training losses.
    }
    \vspace{-4.5mm}
    \label{fig:dual_encoding}
\end{figure*}

\begin{table}[!ht]
    \centering
    \vspace{-2mm}
    \caption{A high-level comparison between codec systems.}
    \vspace{-3mm}
    \resizebox{\columnwidth}{!}{
    \begin{tabular}{llll}
    \hline
        ~ & Semantic Enhancement & Audio Quality & Frame Rate \\ \hline
        Encodec & \ding{55} & Good & 75Hz \\ 
        SpeechTokenizer & \ding{51} (distill) & Good & 50Hz \\ 
        DAC & \ding{55} & \textbf{Great} & 75Hz/50Hz \\ 
        Mimi & \ding{51} (distill) & Good & \textbf{12.5Hz} \\ \hline
        \textbf{DualCodec} & \textbf{\ding{51} (dual encoding)} & \textbf{Great} & \textbf{12.5Hz/25Hz} \\ \hline
    \end{tabular}
    }
    \label{tab:intro}
    \vspace{-2mm}
\end{table}

\noindent A comparison with relevant codec models~\cite{encodec,speechtokenizer,dac,moshi} is presented in Table~\ref{tab:intro}. 
We find that no existing codec model adequately fulfill all three design principles we summarized.
Firstly, 
existing semantic enhancement method distill SSL representation to codec, but we find that the distilled tokens still do not carry enough pronunciation information (evidenced in Sec. \ref{sec:semantic_asr}). 
Also, existing models only attain great audio quality at high frame rates and high bitrates.

We argue that our three summarized design principles can be fulfilled in a single framework: incorporating an explicit semantic-related codec layer, maintaining a low frame rate, and preserving high audio quality. To achieve this, we propose \textbf{DualCodec}, which unifies SSL and waveform representations in a dual encoding framework. 
In this framework, the tokens from the first layer are semantically enhanced directly from SSL feature. 
DualCodec achieves a low frame rate while maintaining exceptional audio quality, with benefits from dual encoding, low-frame-rate configuration, larger codebook sizes, and DAC~\cite{dac} codec innovations. 
We will
open-source the DualCodec in both 12.5Hz and 25Hz settings, as well as the DualCodec-based TTS systems in this work. 
\textit{To the best of our knowledge, this is the first open-source 12.5 Hz low frame-rate codec\footnote{The Mimi codec is the first open-weight 12.5Hz codec, but its data and training code are not available. We train on a public dataset, and have released our model weights and training code (see abstract).}, and the first open-source 12.5Hz TTS}.

\section{Related Works}
\subsection{Semantic Representations}
Self-supervised learning (SSL) have shown strong performance in speech representation learning~\cite{ssl_review, hubert, wavlm, w2v-bert}. 
It is found that lower SSL layers encode speaker-related information whereas higher layers contain rich pronunciation and word meaning information, benefiting tasks such as speech recognition (ASR)~\cite{pasad2021layer_wise_analysis_w2v2_ssl,bertphone,ssl_review}.
Discretizing SSL representations into semantic tokens have also been explored, using k-means~\cite{audioLM2023} or vector quantization (VQ)~\cite{repcodec}. 
However, semantic tokens cannot accurately reconstruct original audios due to a lack of acoustic traits like speaker identity~\cite{audioLM2023}. 

\vspace{-1.5mm}
\subsection{Neural Audio Codecs}
\vspace{-0.5mm}
Vanilla neural audio codecs like Encodec~\cite{encodec} and SoundStream~\cite{soundstream}
consist of an encoder, a residual vector quantization (RVQ) module, and a decoder.
Only waveform is utilized as input.
To serve codec systems better for SLMs, the following three important design decisions have been investigated.

\para{Semantic Enhancement.}
Enhancing the semantic information in audio codec tokens can benefit SLM intelligibility and stability, especially in the RVQ-1 token where the AR generation can accumulate prediction errors~\cite{speechtokenizer}. 
SpeechTokenizer~\cite{speechtokenizer} proposed to enhance RVQ-1 through semantic distillation.
The approach introduces a distillation loss between the
RVQ-1 codebook vector and SSL feature extracted from the same speech input. 
Mimi codec~\cite{moshi} applied the semantic distillation on a 12.5Hz codec.
Concurrent work X-Codec~\cite{ye2024codec_xcodec} concatenates SSL representations with waveform features before RVQ, and jointly optimizes acoustic and semantic reconstruction objectives~\cite{guo2025codec_review}. 
Our dual encoding method differs by encoding RVQ-1 and RVQ-rest tokens from different sources, closely aligning with SLM's AR+NAR paradigm. 
Meanwhile, our work explores low frame rate, audio quality improvement, and multilingual evaluation that are not explored in the concurrent work.

\para{Low Bitrate and Low Frame Rate.}
Vanilla neural audio codecs operate at more than 4kbps bitrate and above 50Hz frame rates\footnote{Frame rate is the number of segments per second in the
codec’s latent representation. Bitrate indicates the data amount per unit of time.}~\cite{encodec,soundstream}.
Some recent efforts focus on designing low bitrate codec systems, benefiting the TTS efficiency by reducing the token length.
WavTokenizer~\cite{wavtokenizer}, SingleCodec~\cite{singleCodec2024} and TS3Codec~\cite{ts3codec}
used a single VQ layer with a high frame rate. 
Mimi codec~\cite{moshi}
reduces frame rate to 12.5Hz while keeping multiple RVQ layers.
The latter low-frame-rate approach has better TTS efficiency by effectively shortening the sequence length in AR part.
However, we still find the previous low-frame-rate codec has unsatisfiable audio quality especially at low bitrates $<$1kbps.

\para{Improving Audio Quality.}
There has been attempts to improve the audio quality over the vanilla Encodec~\cite{encodec} method.
HiFi-Codec~\cite{yang2023hificodec} proposed Grouped RVQ (GRVQ) to improve RVQ information capacity.
Vocos~\cite{vocos} replaced the Encodec decoder with a network to predicting inverse STFT coefficients.
Descript-Audio-Codec (DAC)~\cite{dac} significantly improves audio quality by applying a low-dimensional projection codebook, using a periodic activation function, and applying multi-scale mel loss.
Some recent works explored using Transformer as replacements for CNN modules~\cite{ts3codec,moshi}.
We incorporate the DAC architecture in this work, and leave Transformer codecs as future investigations.

\newcommand{\bmath}[1]{\mathbf{#1}}
\vspace{-3.0mm}
\section{Method}
As shown in Figure \ref{fig:dual_encoding}, our DualCodec system consists of two encoding streams: an SSL encoding stream and a waveform encoding stream. 
\begin{itemize}
    \item The SSL encoding stream captures \textit{semantic-rich} information to the first-layer codec tokens by directly encoding from SSL feature.
    \item The waveform encoding stream encodes and decodes \textit{high-quality} audio with the proven DAC framework.
    \item We apply downsampling to both streams to achieve a \textit{low frame rate}.
\end{itemize}

The two streams are jointly optimized. By using the two encoding streams, we obtain semantic-rich RVQ-1 tokens, with remaining layers (RVQ-rest) focused on the \textit{remaining} acoustic aspects in the waveform feature.
This ``disentanglement" is achieved by subtracting RVQ-1 feature from waveform feature, before obtaining RVQ-rest tokens. 
Finally to decode audio, the RVQ-1 feature is re-summed to the codebook vectors of RVQ-rest.
During TTS training, both encoding streams are required to obtain training tokens.
During TTS inference, only the codec decoder is used to produce audio.

\vspace{-2mm}
\subsection{SSL Encoding} \label{sec:dual-encoding}
\vspace{-1mm}
The SSL encoding stream contains a pretrained SSL model, a ResNet encoder, a downsampler, a vector quantization (VQ) module and a ResNet decoder. 
This architecture is analogous to a VQ-VAE~\cite{vqvae}, inspired by the RepCodec tokenizer~\cite{repcodec} which first applied VQ-VAE to discretizing SSL features.

\firstpara{SSL Model.} 
We use normalized 16th layer w2v-BERT-2.0~\cite{seamless} feature following \cite{maskgct}.
The feature of the deep SSL layer is shown to contain rich pronunciation and word meaning information~\cite{pasad2021layer_wise_analysis_w2v2_ssl,maskgct}.
The model outputs 50Hz feature from 16kHz waveforms with a 600M-parameter Transformer~\cite{transformer} network. 
The SSL model is frozen during training and inference.

\para{Downsampler.}
We downsample the 50Hz SSL feature into our codec's frame rate using simple 1D average pooling, using kenel\_size = stride\_size = downsampling\_factor.
The downsampling\_factor is 2 for 25Hz target frame rate, and is 4 for 12.5Hz.

\para{ResNet Encoder and Decoder.}
These networks are used to process the SSL feature before and after the VQ module.
This allows the VQ tokens to capture more complex semantic patterns.
The decoder mirrors the encoder. Both contain stacked ConvNeXt~\cite{convnext}, which are the latest ResNet~\cite{resnet} variants.
Each network has 13M parameters.
There's no down-sampling or up-sampling operation in these ResNet modules. 

\para{VQ Module.} 
This module discretizes the ResNet Encoder output \(\bmath{Z}_\text{ssl} \in \mathbb{R}^{H \times T}\) into a 1D token sequence \(RVQ\_1 \in \mathbb{Z}^{1 \times T}\), where H is the hidden dimension and T is the feature length.
Following \cite{dac}, $RVQ\_1$ is computed by finding the closest codebook vector to the projected input:
$RVQ\_1 = \arg\min_k ||\ell_2(W_{\text{in}} \bmath{Z}_{ssl}) - \ell_2(e_k)||_2$. Here, $W_\text{in} \in \mathbb{R}^{D \times H}$ is the input projection matrix with $D=8$,$H=1024$, $\ell_2$ is the L2-normalizaton, $e_1, e_2, ..., e_k$ are codebook vectors, $e_k \in \mathbb{R}^{H \times T}$.
We obtain RVQ-1-feature after processing with the ResNet decoder: $RVQ\_1\_feat = \text{ResNet}(e_k)$.

\vspace{-2mm}
\subsection{Waveform Encoding}
\vspace{-1mm}
The waveform encoding stream is inspired by existing neural audio codecs. We adopt the DAC~\cite{dac} architecture, comprising a codec encoder, an RVQ module, and a codec decoder.

\firstpara{Codec encoder and decoder.}
The codec encoder and decoder are CNN networks with snake activation function~\cite{snake-actiFunc}.
The encoder contains a series of strided convolution layers to downsample the waveform into the codec frame rate and extract waveform features.
The decoder mirrors the encoder, replacing strided convolutions with upsampling convolutions.

\para{Frame rate.}
Our model receives 24kHz waveform input.
To output 25Hz frame-rate tokens, the codec encoder uses 4 CNN blocks with strides $(4,5,6,8)$, giving $24000\text{Hz} \div (4 \times 5 \times 6 \times 8) = 25\text{Hz}$.
The 12.5Hz version uses strides $(4,5,6,8,2)$.

\para{RVQ module.} 
This module has $N-1$ layers of VQ.
Each VQ layer quantizes the residual error of the previous layer~\cite{soundstream}. 
The input to this module is the residual between the waveform feature and the $RVQ\_1\_feat$. 
It discretizes into 
RVQ\_rest $\in \mathbb{Z}^{(N-1)\times T}$.
After obtaining the RVQ\_rest tokens, each selected codebook vector $e_k$  is added together with $RVQ\_1\_feat$. 
This continuous feature summarizes SSL encoding and waveform encoding, and is used as input to the codec decoder.
Notably, we employ RVQ dropout~\cite{encodec} during training.
That is, we only use the first $q$ RVQ quantizers each time, where $q \in [0, N-1]$ is randomly chosen.
When $q=0$, only the SSL encoding stream is used, and the model only decodes the RVQ-1 tokens.

\vspace{-1.5mm}
\subsection{Training objective}
\vspace{-1mm}
The dual encoding framework is trained end to end.
It is trained on an added SSL reconstruction loss~\cite{repcodec} on top of the GAN training objective from DAC~\cite{dac}: spectrogram reconstruction loss, quantization loss, and adversarial loss.

\firstpara{SSL reconstruction loss.} 
This is an MSE losss between the reconstructed SSL feature and the input SSL feature. 
Both features are either the 12.5Hz or 25Hz downsampled version. 

\para{Spectrogram reconstruction loss.}
This is a multi-scale Mel Spectrogram loss between the input and reconstructed audio.

\para{Quantization loss}
The codebooks are updated with an L1 loss between quantized and unquantized features.
A commitment loss and the straight-through estimator are also used~\cite{encodec}.

\para{Adversarial loss}
We use Multi-Period Discriminator (MPD) and Multi-Scale STFT Discriminator (MS-STFTD)~\cite{dac,encodec}. An L1 feature matching loss is employed in all intermediate layers between generated and ground truth samples~\cite{dac}.

\section{Experiments}
\subsection{Model training setup}
\vspace{-2mm}
We use the 100K-hour multilingual, 24kHz speech dataset Emilia~\cite{emilia} in Amphion toolkit~\cite{amphion_v0.2}, and 
8 A100 GPUs for training.
Each codec model is trained for 1 epoch (400K steps) with a 3-second segment length.
Each TTS model is trained for 2 epochs (around 600K steps).
There are $N=8$ codebook layers in DualCodec. 
In the following sections, we comprehensively evaluate our codec's semantic content WER, audio reconstruction quality, TTS performance and speed, respectively.

\vspace{-3mm}
\subsection{Semantic content analysis} \label{sec:semantic_asr}
\begin{table}[!ht]
    \centering
    \vspace{-5mm}
    \caption{WER Results of codec-decoded RVQ-1 audio.}
    \vspace{-2mm}
    \resizebox{\columnwidth}{!}{
    \begin{tabular}{llllcc}
    \hline
        \multirow{2}{*}{\textbf{ID}} & \multirow{2}{*}{\textbf{Rate}} & \multirow{2}{*}{\textbf{Method}} & \textbf{RVQ-1} & \textbf{EN } & \textbf{ZH }  \\
        &&&\textbf{Config} & \textbf{WER(\%)$\downarrow$} & \textbf{WER(\%)$\downarrow$} \\ \hline
        A1 & - & Ground-Truth & - & 2.13 & 1.25 \\ 
        A2 & 50Hz & SpeechTokenizer & 1024 EMA & 14.9 & 83.2 \\ \hline
        B1 & 25Hz & DAC & 1024 Proj & 55.4 & 46.4 \\ 
        B2 & 25Hz & w/ Distill & 1024 Proj & 28.4 & 26.4 \\ 
        B3 & 25Hz & w/ Dual encoding & 1024 Proj & \textbf{5.59} & \textbf{6.52} \\ \hline 
        C1 & 25Hz & DAC & 16384 Proj & 31.8 & 21.0 \\ 
        C2 & 25Hz & w/ Distill & 16384 Proj & 17.8 & 14.4 \\ 
        C3 & 25Hz & w/ Dual encoding & 16384 Proj & \textbf{2.98} & \textbf{2.91} \\ \hline
        D1 & 12.5Hz & w/ Dual encoding & 16384 Proj & 6.94 & 6.36 \\ \hline
    \end{tabular}
    }
    \vspace{-3mm}
    \label{tab:semantic_framerate}
\end{table}

\begin{table*}[htbp]
    \centering
    \vspace{-3mm}
    \caption{Audio reconstruction performance of neural audio codecs around 75token/s and 0.75kbps bitrate on LibriSpeech-test-clean.}
    \vspace{-3mm}
    \resizebox{\textwidth}{!}{
    \begin{tabular}{llllllllllll}
    \toprule
        \textbf{ID} & \textbf{System} \textit{(RVQ-1 size, RVQ-rest size)} & \textbf{Params} & \textbf{Bit(kbps)} & \textbf{Tok/s} & \textbf{\#VQ} & \textbf{PESQ\_nb$\uparrow$} & \textbf{PESQ\_wb$\uparrow$} & \textbf{STOI$\uparrow$} & \textbf{MCD$\downarrow$} & \textbf{UTMOS$\uparrow$} & \textbf{MUSHRA$\uparrow$}\\ \midrule
        E1 & DAC-official 75Hz & 74M & 0.75 & 75 & 1 & 1.46 & 1.18 & 0.75 & 6.00 & 1.32 & 26.0  \\ 
        E1 & Encodec 75Hz & 15M & 1.5 & 150 & 2 & 1.92 & 1.54 & 0.84 & 4.30 & 1.55 & 36.2 \\
        E3 & SpeechTokenizer 50Hz & 103M & 1.0 & 100 & 2 & 1.42 & 1.15 & 0.70 & 6.94 & 1.81 & 35.9 \\ 
        E4 & WavTokenizer-large 75Hz & 71M & 0.90 & 75 & 1 & \textbf{2.54} & \textbf{2.05} & \textbf{0.89} & \textbf{3.99} & \textbf{3.87} & \textbf{81.0}  \\ 
        E5 & Mimi 12.5Hz & 78M & 0.83 & 75 & 6 & 2.51 & 1.99 & \text{0.89} & 4.13 & \text{3.43} & 72.8 \\ 
        \hline
        F1 & DAC-repro 25Hz \textit{(1024+1024)} & 74M & 0.75 & 75 & 3 & 2.58 & 2.06 & 0.89 & 3.93 & 3.29 & 68.8  \\ 
        F2 & DAC-repro 12.5Hz \textit{(1024+1024)} & 69M & 0.75 & 75 & 6 & \textbf{2.88} & \textbf{2.33} & \textbf{0.91} & \textbf{3.70} & \textbf{3.87} & \textbf{81.8} \\ 
        \hline
        G1 & DualCodec 25Hz \textit{(1024+1024)} & 100M & 0.75 & 75 & 3 & 2.64 & 2.07 & 0.90 & 3.99 & 3.86 & 79.5  \\ 
        G2 & DualCodec 25Hz \textit{(16384+1024)} & 100M & 0.85 & 75 & 3 & 2.92 & 2.32 & 0.91 & 3.61 & 4.08 & 86.2 \\ 
        G3 & DualCodec 12.5Hz \textit{(1024+1024)} & 95M & 0.75 & 75 & 6 & 2.89 & 2.30 & 0.91 & 3.61 & 3.94 & 83.5\\
        G4 & DualCodec 12.5Hz \textit{(16384+1024)} & 95M & 0.80 & 75 & 6 & 2.94 & 2.33 & 0.91 & 3.65 & 4.04 & 85.2 \\ 
        G5 & DualCodec 12.5Hz \textit{(16384+4096)} & 95M & 0.93 & 75 & 6 & \textbf{3.11} & \textbf{2.54} & \textbf{0.92} & \textbf{3.33} & \textbf{4.11} & \textbf{88.2} \\ \hline
    \end{tabular}
    }
    \vspace{-4mm}
    \label{tab:codec_performance}
\end{table*}
\firstpara{Metrics.} We evaluate the semantic preservation of the RVQ-1 tokens by reporting the ASR Word Error Rate (WER) on the codec-decoded audio, using only RVQ-1. The evaluations leverage Whisper-large-v3 for English (EN) and Paraformer-zh for Chinese (ZH), tested on the Seed-TTS-Eval~\cite{seedtts} dataset. 
The results are summarized in Table~\ref{tab:semantic_framerate}.

\para{Group A baselines.}
Group A (A1 and A2) models are ground-truth and official SpeechTokenizer checkpoint, respectively.
The A2 SpeechTokenizer model has semantic distillation and 1024 EMA codebooks.
While it is trained on English-only data, we report its Chinese performance and find that its RVQ-1 has extremely high Chinese WER. 
Listening to samples, we suggest that the model's RVQ-1 tokens, although having semantic distillation, lacks pitch information. Pitch is critical for understanding Chinese.

\para{Effect of Dual Encoding.} Without semantic enhancement, the DAC framework at 25Hz (B1) yields high WERs of 55.4 (EN) and 46.4 (ZH). 
We then retrain with added semantic distillation loss, this (B2) reduces WERs to 28.4 and 26.4.
Using dual encoding under the same RVQ setting (B3) further improves performance, achieving WERs of 5.59 (EN) and 6.52 (ZH). These results highlight the effectiveness of dual encoding in significantly enhancing semantic preservation.

\para{Effect of Larger Codebooks.} Increasing the RVQ-1 codebook size to 16384 brings additional improvements. With dual encoding(C3), the WERs drop to 2.98 (EN) and 2.91 (ZH), closely approaching the ground truth. 
Meanwhile, at half the frame rate at 12.5Hz, the dual encoding configuration (D1) achieves competitive results, with WERs of 6.94 (EN) and 6.36 (ZH).

\vspace{-2mm}
\subsection{Audio quality analysis}
\vspace{-1mm}
\firstpara{Metrics.} In this section, we report the audio reconstruction quality of DualCodec, using the full Librispeech-test-clean \cite{librispeech} data under a consistent 75 tokens per second and around 0.75kbps low bitrate.
Metrics include the Perceptual Evaluation of Speech Quality (PESQ) \cite{pesq} (both the 8kHz narrow-band PESQ\_nb, and 16kHz wide-band PESQ\_wb), Short Term Objective Intelligibility (STOI) \cite{stoi}, Mel Cepstral Distortion (MCD) \cite{mcd}. 
We also evaluate on the reference-free neural MOS predictor UTMOS~\cite{saeki2022utmos}, a metric that highly correlates with human preferences~\cite{saeki2022utmos,ts3codec}.
The subjective test is the MUltiple Stimuli with Hidden Reference and Anchor (MUSHRA)~\cite{webmushra}.
We conduct the test with 8 participants rating 15 sets of audios reconstructions sampled from the same test set. 
The baselines include official DAC~\cite{dac}, Encodec~\cite{encodec}, SpeechTokenizer\cite{speechtokenizer}, WavTokenizer~\cite{wavtokenizer}, and Mimi~\cite{moshi}.
The number of \textit{trainable} codec parameters are also indicated.
Results are presented in Table~\ref{tab:codec_performance}.

\para{Baseline Systems.} Among the baseline systems (group E), WavTokenizer-large stands out as the strongest performer across both objective and subjective metrics. 
Mimi also performs well, slightly trailing behind WavTokenizer-large.
The remaining baselines, originally targeting high bitrates, has significantly lower audio quality in this low-bitrate scenario.

\para{Reproduced DAC.} Group F focuses on our retrained DAC codec modified for 25Hz and 12.5Hz frame rates.
At the same 0.75 kbps bitrate, the 12.5 Hz DAC obtain much higher performance than 25Hz in every metric. 
This suggests that operating at a lower frame rate with more quantization layers is more effective than a larger frame rate with less quantization layers.
Interestingly, our 12.5Hz DAC variant outperforms all baseline models, suggesting \textit{the effectiveness of the DAC framework especially at lower frame rates}.

\para{DualCodec.} Group G highlights DualCodec's performance under ablated settings. 

\para{1. Effect of Dual Encoding.} Models labeled G1 and G3 have same RVQ codebook settings as Group F models, allowing us to ablate on the impact of dual encoding.
Comparing G1 (DualCodec 25Hz) with F1 (DAC-repro 25Hz), G1 achieves similar performance across most objective metrics but demonstrates a significant improvement in UTMOS (3.86 vs. 3.29) and MUSHRA (79.5 vs. 68.8), indicating DualCodec's noticeable enhancement in perceptual audio quality.
Comparing G3 (DualCodec 12.5Hz) with F2 (DAC-repro 12.5Hz) shows a similar trend where DualCodec has higher perceptual metrics UTMOS (3.94 vs. 3.87) and MUSHRA (83.5 vs. 81.8).
We conclude that \textit{DualCodec’s additional semantic encoding stream can enhance the perceptual audio quality}.

\para{2. Effect of Larger Codebooks. }We examine the impact of increasing the RVQ codebook sizes in DualCodec, which slightly increases the bitrate but maintains a consistent token rate of 75 tokens/s. 
Model G2, with an increased 16384 codebooks in RVQ-1, shows marked improvements over G1 in every metric (UTMOS: 4.08 vs. 3.86, MUSHRA: 86.2 vs. 79.5).
The trend continues in models G4 and G5 in the 12.5Hz scenario. 

In Model G5, we experiment with 
further increasing the codebook size of each RVQ-rest layer to 4096, a similar size as used in WavTokenizer~\cite{wavtokenizer}. The model achieves the best overall performance among all systems, with PESQ\_nb=3.11, STOI=0.92, UTMOS=4.11, and the highest MUSHRA score (94.8). 
This result highlights that 
\textit{increasing the codebook size while leveraging lower frame rates can significantly enhance low-bitrate audio quality. }

\vspace{-1.5mm}
\subsection{TTS analysis}
\begin{table}[!ht]
    \centering
    \vspace{-5mm}
    \caption{Codec-based TTS performance on Seed-TTS-Eval.}
    \vspace{-3mm}
    \resizebox{\columnwidth}{!}{
    \begin{tabular}{llccccc}
    \toprule
        \multirow{2}{*}{\textbf{TTS}} & \multirow{2}{*}{\textbf{Codec}} & \textbf{EN} & \textbf{EN} & \textbf{ZH} & \textbf{ZH} & \multirow{2}{*}{\textbf{RTF$\downarrow$}}  \\ 
        && \textbf{WER$\downarrow$} & \textbf{SIM$\uparrow$} & \textbf{WER$\downarrow$} & \textbf{SIM$\uparrow$} & \\ \midrule
        GT & - & 2.13 & 0.73 & 1.25 & 0.75 & -  \\ \hline
        \multirow{4}{*}{\shortstack[l]{VALL-E}} & SpeechTokenizer & 15.4 & 0.47 & 21.5 & 0.55 & 0.76  \\ 
        ~ & Mimi & 8.16 & 0.48 & 10.5 & 0.55 & \textbf{0.16}  \\ 
        ~ & DualCodec 25Hz & \textbf{3.40} & \textbf{0.57} & \textbf{2.49} & \textbf{0.67} & 0.30  \\ 
        ~ & DualCodec 12.5Hz & 4.40 & 0.54 & 4.90 & 0.65 & \textbf{0.16}  \\ \hline
        \multirow{4}{*}{\shortstack[l]{AR +\\SoundStorm}} & SpeechTokenizer & 11.3 & 0.50 & 46.3 & 0.57 & 1.18  \\ 
        ~ & Mimi & 9.09 & 0.50 & 39.4 & 0.57 & \textbf{0.34}  \\ 
        ~ & DualCodec 25Hz & \textbf{3.56} & \textbf{0.67} & \textbf{2.93} & \textbf{0.75} & 0.66  \\ 
        ~ & DualCodec 12.5Hz & 4.93 & 0.59 & 4.72 & 0.69 & \textbf{0.34} \\ \hline
    \end{tabular}
    }
    \vspace{-3mm}
    \label{tab:slm_performance}
\end{table}

\firstpara{Metrics.} 
We adopt VALL-E~\cite{valle} and AR+SoundStorm~\cite{soundstorm2023,maskgct} as TTS systems and train each TTS with different codec systems.
VALL-E has 270M parameters in AR, 400M in NAR.
AR+SoundStorm has 800M in AR, 300M in NAR.
For DualCodec, models G2 and G5 in Table \ref{tab:codec_performance} are used for 25Hz and 12.5Hz, respectively. For all codecs we use 8 RVQ codebooks.
We report the WER and speaker similarity SIM-O (SIM) on Seed-TTS-Eval benchmark~\cite{seedtts}. We report the real-time-factor (RTF) tested on an A100 GPU which correlates to the inference speed. 
Results are shown in Table \ref{tab:slm_performance}.

\para{Performance Comparisons.}
Table \ref{tab:slm_performance} demonstrates that DualCodec outperforms SpeechTokenizer and Mimi baselines in both TTS. 
We attribute the superior WER performance to DualCodec’s more accurate semantic content, and the SIM to DualCodec's better codec reconstruction quality. 
The AR+SoundStorm TTS paired with DualCodec 25Hz achieves the best performance, followed by DualCodec 12.5Hz.
When comparing RTF, the trade-off between quality and inference speed becomes apparent: 
DualCodec 25Hz consistently achieves lower WER and higher SIM scores, making it the ideal choice for tasks prioritizing accuracy and similarity. 
On the other hand, DualCodec 12.5Hz provides much faster inference at the cost of acceptable degrees of performance decrease. 
We also notice that Mimi and SpeechTokenizer have excessively large Chinese WERs in AR+SoundStorm.
We suggest this is due to a lack of RVQ-1 semantic pitch information, which becomes more notable in AR+SoundStorm TTS because its NAR does not have text prompting for extracting the necessary pitch.

\vspace{-2.5mm}
\section{Conclusion}
\vspace{-1.5mm}
We introduced DualCodec, a low-frame-rate, semantically-enhanced neural audio codec designed for efficient speech generation. By leveraging dual encoding, low frame rates and larger codebooks, DualCodec improves semantic accuracy, audio reconstruction quality, TTS performance and efficiency. 
A limitation in this work is the gap between the 12.5Hz DualCodec-based TTS and the 25Hz ones.
Future works could explore methods to further increase the 12.5Hz codec's semantic accuracy and quality upper-bound.
DualCodec also has the potential to be used in real-time multimodal LLMs, since it allows both efficient audio encoding and high quality audio reconstruction.

\vspace{-2mm}
\section{Acknowledgements}
The work is partially supported by 
the NSFC under Grant 62376237; 
the Shenzhen Science and Technology Program ZDSYS20230626091302006; 
2023 Shenzhen Stability Science Program; Program for Guangdong Introducing Innovative and Entrepreneurial Teams 2023ZT10X044.

\section{Appendix}
\appendix
\section{Detailed Architecture of TTS Systems} \label{sec:appendix_slm_arch}
\subsection{VALL-E} 
\begin{figure}[htbp]
    \centering
    \includegraphics[width=\columnwidth]{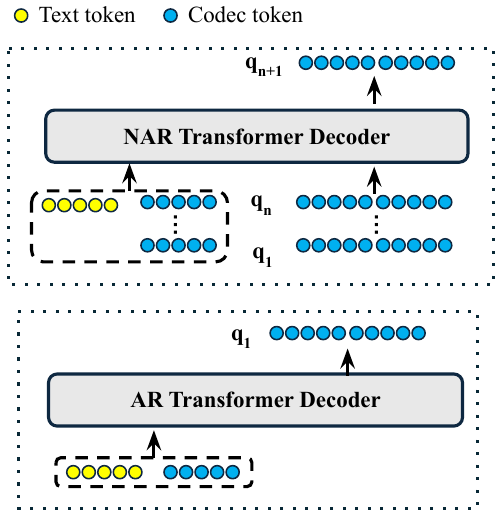}
    \caption{Architecture of VALL-E TTS.}
    \label{fig:valle}
\end{figure}
The architecture of VALL-E~\cite{valle} TTS is shown in Figure \ref{fig:valle}~\cite{codecinvestigation}.
The model can do zero-shot TTS: Given an unseen prompt speech and a text sequence to generate, the model can imitate the speaker speaking the target text.

There are two stages in VALL-E, an AR stage and an NAR stage.
First, the AR model takes as input 
a concatenation of text tokens and 
the RVQ-1 tokens extracted from prompt speech.
The text tokens are encoded with Byte-Pair-Encoding (BPE) from Whisper~\cite{whisper}.
The target text is concatenated after the prompt text during inference.
It then predicts the RVQ-1 codec tokens for the target speech in an autoregressive (AR) manner. 

Following this, the NAR model predicts the remaining codec tokens (RVQ-rest) layer-by-layer. 
For each subsequent layer, the NAR model predicts based on all previously predicted layers, combining them with the text token sequence and prompt codec tokens from all quantization layers, operating in parallel.
Both models use Transformer architecture with relative positional encoding, but the AR model predicts causally, whereas the NAR model predicts tokens in parallel.
The two stages are trained separately, using a cross-entropy loss with teacher forcing.

\subsection{AR+SoundStorm}
\begin{figure}
    \centering
    \includegraphics[width=\columnwidth]{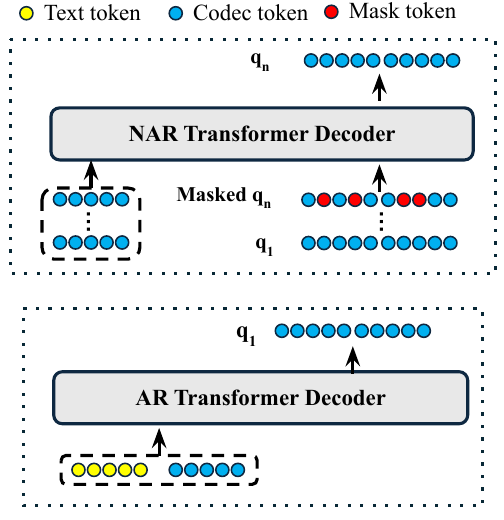}
    \caption{Architecture of AR+SoundStorm TTS.}
    \label{fig:soundstorm}
\end{figure}

The architecture of the AR+SoundStorm~\cite{soundstorm2023} TTS is shown in Figure \ref{fig:soundstorm}.
Its AR stage has the same architecture as in VALL-E and predicts RVQ-1 tokens.
Different from VALL-E, the NAR stage uses a masked token prediction strategy to predict RVQ-rest tokens layer by layer. 
The method is inspired by MaskGIT~\cite{maskGIT2022} from the image domain, and has shown better audio quality than VALL-E at a cost of slower, multi-step inference.

During training, 
a target RVQ layer $n \in (1, N]$ is randomly chosen.
Given previously predicted codec tokens, the model randomly masks a subset of tokens in the RVQ-$n$ layer and predicts them in parallel. 
The masked tokens are replaced with a special \texttt{[MASK]} token.
We set the probability that each token is masked be $p(t) = sin(\frac{\pi t }{2})$, in which a timestep variable $t \in [0, 1)$ is sampled uniformly in each training step. 
When $t=0$, all tokens are unmasked; when $t=1$, all tokens are masked.

During inference,
the masked tokens are iteratively refined through multiple decoding steps based on confidence scores, progressively filling in all layers of codec tokens.
For example, assuming the decoding steps for predicting RVQ-2 is $S$, 
the decoding starts with a fully masked RVQ-2 sequence \( \mathbf{X}_T \) with RVQ-1 as the condition. 
In each step $i \in [1,S]$, the model predicts all masked tokens, but only a total of $N p(\frac{i}{S})$ tokens are retained, where $N$ is the token sequence length.
The decoding step $S$ is an important hyper-parameter that balances the inference speed and fidelity. 
We use [20,10,1,1,1,1,1] as the $S$ for the seven RVQ-rest layers~\cite{maskgct}.

\section{The Inference Speed of DualCodec} \label{appendix:codec_speed}
\begin{table}[!ht]
    \centering
    \caption{The parameter count and the real-time factor (RTF) to encode and decode different neural audio codecs, tested on an A100 GPU with a batch size of 1 on LibriSpeech-test-clean.}
    \resizebox{\columnwidth}{!}{
    \begin{tabular}{lllll}
    \toprule
        \multirow{2}{*}{Model} & \multicolumn{2}{c}{Encode} & \multicolumn{2}{c}{Decode} \\ 
        \cmidrule(lr){2-3} \cmidrule(lr){4-5}
        & \#Params & RTF$\downarrow$ & \#Params & RTF$\downarrow$ \\ \midrule
        DAC-official & 22M & 0.005 & 52M & 0.004 \\ 
        Encodec & 7.4M & 0.006 & 7.4M & 0.004 \\ 
        SpeechTokenizer & 68M & 0.006 & 35M & 0.004 \\ 
        WavTokenizer-large & 8.8M & 0.006 & 62M & \textbf{0.002} \\ 
        Mimi & 38M & 0.008 & 40M & 0.004 \\ 
        DAC-repro 25hz & 22M & \textbf{0.003} & 52M & 0.002 \\ 
        DAC-repro 12hz & 16M & 0.004 & 53M & \textbf{0.002} \\ 
        DualCodec 25hz & 628M & 0.015 & 53M & \textbf{0.002} \\ 
        DualCodec 12hz & 622M & 0.015 & 53M & 0.003 \\ \bottomrule
    \end{tabular}
    }
    \label{tab:codec_speed}
\end{table}

\firstpara{Metrics.}
To provide a comprehensive evaluation on the inference speed of DualCodec, we evaluate the Real-Time Factor (RTF) and report the parameter counts for encoding and decoding across various codec systems. 
RTF measures the ratio of processing time to audio duration. 
In Table~\ref{tab:codec_speed}, we present the RTF results of different codec systems.
We report the amortized RTF over LibriSpeech-test-clean measured on an A100 GPU with a batch size of 1.

\para{Results.}
From Table~\ref{tab:codec_speed}, we observe that DualCodec's RTFs, while not the most competitive among the evaluated systems, remain very acceptable given its design and capabilities. For instance, DualCodec 25Hz achieves an encoding RTF of 0.015 and a decoding RTF of 0.002, and DualCodec 12Hz attains an encoding RTF of 0.015 and a decoding RTF of 0.003. 
Importantly, the decoding phase of DualCodec remains highly efficient, with RTF values on par with or better than most baselines.
This ensures the DualCodec-based TTS systems remains lightweight and fast during deployment.

\section{More Open-Source DualCodec TTS Systems} \label{sec:more_dualcodec_TTS}
\begin{table*}[h!]
    \centering
    \caption{A summary of speech language model performance on Seed-TTS-Eval. We add ``DualCodec-'' prefix for DualCodec-based models. We are open-sourcing DualCodec-VALLE, DualCodec-MaskGCT, and DualCodec-FlattenedAR. See the demo page link for more details. }
    \resizebox{\textwidth}{!}{
    \begin{tabular}{llllll}
    \toprule
        ~ & \textbf{Tokenizer} & \textbf{ZH WER$\downarrow$} & \textbf{ZH SIM-O$\uparrow$} & \textbf{EN WER$\downarrow$} & \textbf{EN SIM-O$\uparrow$}  \\ \midrule
        Ground-Truth & ~ & 1.25 & 0.75 & 2.13 & 0.73  \\ 
        CosyVoice~\cite{cosyvoice} & S3Tokenizer+flow matching 50Hz Mel & 3.10 & 0.75 & 3.39 & 0.64  \\ 
        FireRedTTS~\cite{fireredtts} & In-house 25Hz & \textbf{1.51} & \text{0.63} & 3.82 & 0.46  \\ 
        AR+SoundStorm~\cite{maskgct} & RepCodec+DAC 50Hz & 2.97 & 0.73 & 3.41 & 0.66  \\ 
        VoiceCraft~\cite{voicecraft} & Encodec 50Hz & - & - & 7.57 & 0.47  \\ 
        MaskGCT~\cite{maskgct} & RepCodec+DAC 50Hz & 2.27 & \textbf{0.77} & \textbf{2.62} & \textbf{0.72}  \\ \midrule 
        DualCodec-VALLE 25Hz & DualCodec 25Hz & 2.49 & 0.67 & 3.40 & 0.57  \\ 
        DualCodec-VALLE 12.5Hz & DualCodec 12.5Hz & 4.90 & 0.65 & 4.40 & 0.54  \\ 
        DualCodec-AR+SoundStorm 25Hz & DualCodec 25Hz & 2.93 & \textbf{0.75} & 3.56 & \textbf{0.67}  \\ 
        DualCodec-AR+SoundStorm 12.5Hz & DualCodec 12.5Hz & 4.72 & 0.69 & 4.93 & 0.59  \\ 
        DualCodec-FlattenedAR & DualCodec 12.5Hz(layer1:4) & 4.68 & 0.56 & \textbf{3.18} & 0.50 \\ 
        DualCodec-MaskGCT 25Hz & DualCodec 25Hz & \textbf{1.85} & 0.75 & 4.18 & 0.68  \\ 
        DualCodec-MaskGCT 12.5Hz & DualCodec 12.5Hz & 3.32 & 0.69 & 5.07 & 0.59  \\ \bottomrule
    \end{tabular}
    }
    \label{tab:appendix_TTS_performance}
\end{table*}

We are open sourcing more TTS systems based on DualCodec, trained on the 100K-hour Emilia~\cite{emilia} dataset.
Apart from the VALL-E and AR+SoundStorm TTS, we also demonstrate DualCodec's effectiveness for an AR-only FlattenAR architecture, and the MaskGCT~\cite{maskgct} architecture.

\begin{figure}
    \centering
    \includegraphics[width=\columnwidth]{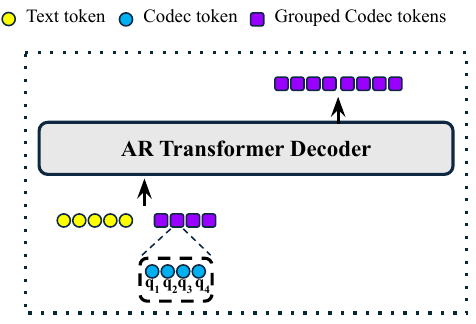}
    \caption{The architecture of a simple FlattenedAR TTS. It models a flattened sequence of four layer RVQ tokens from DualCodec 12.5Hz.}
    \label{fig:flattened_ar_arch}
\end{figure}

\firstpara{Architecture of FlattenAR TTS.}
The architecture of a simple FlattenAR TTS is presented in Figure \ref{fig:flattened_ar_arch}.
The model uses a single AR Transformer to model a flattened sequence of 4-layer RVQ tokens from DualCodec 12.5Hz. This makes the resulting speech token rate 50Hz. 
Compared to other two-stage TTS systems, the model stands out for its simplicity and similarity to an LLM.
Our DualCodec 12.5Hz has advantage in this architecture because it is low-frame-rate and has high audio quality.
While using DualCodec 25Hz or other higher-frame-rate codecs are possible, the token rate ($4 \times 25$Hz=100Hz) is too high and can dramatically slow down the model.

\para{Architecture of DualCodec-based MaskGCT.}
\begin{figure}
    \centering
    \includegraphics[width=\columnwidth]{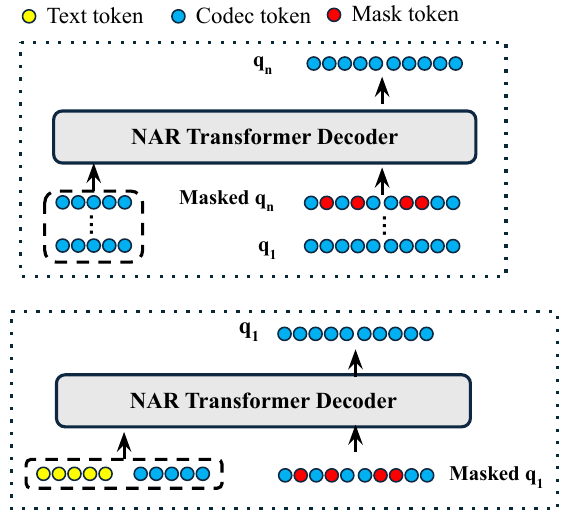}
    \caption{Architecture of MaskGCT TTS. Compared to the AR+SoundStorm TTS, its first stage instead uses the masked token prediction paradigm.}
    \label{fig:maskgct_arch}
\end{figure}
The architecture of the DualCodec-based MaskGCT~\cite{maskgct} TTS is shown in Figure \ref{fig:maskgct_arch}. 
Compared to AR+SoundStorm, MaskGCT uses masked token prediction step 
in its first stage. 

\para{Benchmarking on SeedTTS-Eval.}
In table \ref{tab:appendix_TTS_performance}, we summarize the performance of various TTS on the Seed-TTS-Eval dataset.

Comparing DualCodec-based speech language models with others, we see that DualCodec-based TTS models are the only solution offering 12.5Hz low frame rate, and the 25Hz TTS are competitive to 50Hz baselines.
Some baselines use RepCodec or S3Tokenizer to extract its first-stage semantic tokens, and DAC for its second-stage tokens. 
Our DualCodec-based solution has does not need separate training of semantic and acoustic tokenizers, avoiding the need to re-align semantic and acoustic (audio codec) token frame-rates, and removing the complexity of duplicate information in semantic and acoustic tokens.

\bibliographystyle{IEEEtran}
\bibliography{mybib}

\end{document}